\documentstyle[12pt]{article}

\begin{document}
\thispagestyle{empty}

\begin{center}
               RUSSIAN GRAVITATIONAL SOCIETY\\
               INSTITUTE OF METROLOGICAL SERVICE \\
               CENTER OF GRAVITATION AND FUNDAMENTAL METROLOGY\\

\end{center}
\vskip 4ex
\begin{flushright}
                                         RGS-CSVR-005/95
                                         \\ gr-qc/9508050

\end{flushright}
\vskip 15mm

\begin{center}
{\large\bf Some Problems of Topology Change Description in the Theory
  of Space-Time } \\

\vskip 5mm
{\bf M.Yu. Konstantinov }\\
\vskip 5mm
     {\em VNIIMS, 3-1 M. Uljanovoj str., Moscow, 117313, Russia}\\
     e-mail: konst@cvsi.rc.ac.ru \\
\end{center}
\vskip 10mm

\begin{abstract}
  The problem of topology change description in gravitation theory is analized
in detailes. It is pointed out that in standard four-dimensional theories the
topology of space may be considered as a particular case of boundary conditions
(or constraints). Therefore, the possible changes of space topology in
(3+1)-dimensions do not admit dynamical description nor in classical nor in
quantum theories and the statements about dynamical supressing of topology
change have no sence. In the framework of multidimensional theories the space
(and space-time) may be considered as the embedded manifolds. It give the real
posibilities for the dynamical description of the topology of space or
space-time.
\end{abstract}

\vskip 10mm

\vskip 30mm

\centerline{Moscow 1995}
\pagebreak


\section{Introduction}

The assumption that topology of 3-space may change dynamically or undergo
quantum fluctuations was stated for the first time by Wheeler in connection
with his geometrodynamics program~\cite{wheeler}. According to Wheeler
observable properties of matter and fields must be explained by geometrical
and topological properties of space-time and the dynamics of matter and
fields configuration are the result of the dynamics of space-time geometry
and topology. This assumption and its motivations were discussed by many
authors~\cite{brill}-\cite{borde}. It was shown that the properties of
matter and fields is indeed closely connected with the topology of
space-time~\cite{denardo}-\cite{gonchar}. Some analogies between the
topology of space-time and the properties of matter fields were also found
recently~\cite{smolin}. The considerable progress was achieved in the
investigation of in the investigation of the ''existence'' and properties of
the topology change models~\cite{geroch}, \cite{lee}, \cite{sorkin}-\cite
{selectrules}, where the well known results of differential geometry and
topology may be used. The most essential results in this areas are the
theorems about global hyperbolicity and Cauchy problem in general relativity~%
\cite{hawkellis}-\cite{Vayes}, the Geroch theorem~\cite{geroch} and it's
generalization~\cite{lee}. The theorems about global hyperbolicity state
that globally hyperbolic space-time model (i.e. the space-time model which
may be considered as a solution of the Cauchy problem) must have topology of
the direct product ${\bf M}^3\times {\bf R}^1$ of 3-dimensional space ${\bf M%
}^3$ and the real line ${\bf R}^1$. The theorem of Geroch~\cite{geroch},
\cite{lee} may be considered as a supplement to the global hyperbolicity
theorems. It states that space-time model with different topologies of space
sections must be singular or contain closed time-like curves. These
statements jointly establish the impossibility of dynamical consideration
(i.e. as a solution of some Cauchy problem) of classical topology changes
processes in the framework of four-dimensional theory but they do not forbid
to consider topology change models as solutions of some boundary problem. As
all experiments yields only local data, we have no {\it a priori} basis for
excluding from consideration the space-time models which solve some boundary
problem, but the physical meaning of such models is unclear.

Some additional restrictions on the topology change models were imposed also
in a number of papers~\cite{sorkin}, \cite{horowits}, \cite{selectrules} but
the most results of such type were obtained under different additional
assumptions which concerns the global properties of both space-time itself
and some additional structures. The realization of such assumptions in the
real space-time do not follow from observations or from some fundamental
physical principles which are essentially local. For instance, in the recent
proposal of so called ''selection rules for topology changes'' by Gibbons
and Hawking~\cite{selectrules} the existence of global spin structure is
supposed. However the existence of any global structure
does not follows directly from observations. If we requier the existence
only local spin structure than the more general class of four-manifolds than
it claimed in~\cite{selectrules} and hence more general topological changes
models will be admitted.

The particular models of topology changes were also discussed by several
authors~\cite{borde}, \cite{carlip}-\cite{kirits2}, \cite{konstmel86}, \cite
{konstmel88}. Unfortunately, all such models are made ''by hand'' and there
is no topology change models ware obtained as a solution of some dynamical
or boundary problem. Moreover there is almost no progress in the
constructive description of topology change: up to now there is no real (or
toy) theory which describe the possible changes of space topology. All
existing attempts to describe topology change or topological fluctuation are
phenomenological. It is often supposed that topology changes are pure
quantum phenomena~\cite{brill}, \cite{hawk78}, \cite{foam}, \cite
{selectrules}, but there is no real progress in their quantum description
also.

Such situation indicate that the difficulties which arise in connection with
the problem of topology change description are of principal nature. It is
indeed the case because to describe the manifold structure the following
data must be given: ({\it i}) the finite or countable set of coordinate maps
and the order of their junction; ({\it ii}) the set of functions which
connect the coordinate systems of different maps in their intersections~\cite
{milnor}, \cite{kobayashi}. These data must be given before the solution of
any equations. They may be considered as constraints or as a part of
boundary conditions which are given ''by hand''. Therefore to describe the
changes of space or space-time topology the number of the coordinate maps,
the order of their junction and the junction conditions must be converted
from the class of constraints or boundary data into the class of dynamical
variables. This problem does not trivial because any change of the number of
coordinate maps and identification or re-identification of points of
space-time must induce the redefining of the space of functions on it.
However, the formalism of the current field theory does not contain any
tools which make possible to change the number of coordinate maps and to
perform identification of space or space-time points including the
redefinition of the function space. By this reason, the Hawking's proposal
to include degenerate metrics into path integrals for description of the
topological fluctuations~\cite{foam} does not solve the problem because the
values of functions along degenerate space path do not identified
automatically. The same is true for the so-called 'minimalist wormhole
model' which was supposed recently by Smolin~\cite{smolin}.

The main goal of this paper is to analyze formally several aspects of
topological change description, consider several different attempts to the
solution of this problem and discuss some perspectives. For this purpose in
the next section we discuss the role of space-time topology in the field
theory. In particular, using the results of our previous papers~\cite
{konst83n4}, \cite{padova1} it will be formalized the above statement that
in the current classical field theory topology of space-time is a
constraint. Some possibilities of the topology change consideration in the
framework of $(3+1)$-dimensional theory will be discussed in the section 3.
The rest part of paper contains some discussion and speculations about
different possibilities of the topology change description in
multidimensional space-time theory.

\section{Topology of space-time in current field theory}

Consider the action integral of the some field theory (both classical and
quantum in its path integral form) in four-dimensional space-time in the
following general form
\begin{equation}
\label{action}S=\int\limits_{{\bf M}} L(\Phi _A,\Phi _A,_\alpha )d^4\sigma ,
\end{equation}
where $L(\Phi _A,\Phi _A,_\alpha )$ is the Lagrangian which is depend from
the fields potentials $\Phi _A$ and their derivatives, $A$ is the cumulative
index and $d^4\sigma $ is the invariant volume element which in the local
coordinates $\left\{ x^\alpha ,\alpha =0,...,3\right\} $ has the form
\begin{equation}
\label{vol}d^{4}\sigma =\sqrt{-g}d^{4}x=\sqrt{-g}dx^{0} \Lambda ...\Lambda
dx^3
\end{equation}
where $\Lambda $ is the exterior product of differential forms, $%
g=\det\left\| g_{\alpha \beta }\right\| $ and $g_{\alpha \beta }$ is the
metric tensor of Lorentzian signature $diag(+, -, -, -)$ on ${\bf M}$.

In the equality~(\ref{action}) the integration are carried out over the full
manifold ${\bf M} $, so both action $S$ and corresponding Feynman amplitude $%
\exp\left\{iS/h\right\} $ are the functionals of both field variables $%
\Phi_A $ and the manifold itself.

To formalize the action dependence from the manifold structure, let us
consider some atlas ${\bf U}=\left\{ {\bf V}_k\right\} $ of ${\ }{\bf M}$,
i.e. finite or countable covering of ${\bf M}$ by coordinate maps ${\bf V}_k$
which are diffeomorphic to unit cube ${\bf D}^4$ of the Euclidean space $%
{\bf R}^4:{\bf M}=\bigcup\limits_{k\in {\bf P}}{\bf V}_k$, where ${\bf V}%
_k\sim {\bf D}^4$ and ${\bf P}\subset {\bf N}$ is a subset of the set ${\bf N%
}$ of natural numbers, which numerate the elements of the covering ${\bf U}$%
. Let $\left\{ x_k^\alpha \right\} $ are the local coordinates in the region
${\bf V}_k$ and $\left\{ x_{i_0...i_l}^\alpha \right\} $ are some local
coordinates in the intersection ${\bf V}_{i_0}\cap ...\cap {\bf V}_{i_l}$
(which is also diffeomorphic to ${\bf D}^4$). In any intersection ${\bf V}%
_{i_0}\cap ...\cap {\bf V}_{i_l}$ the field potentials $\Phi _A$ must
satisfy to the natural consistency conditions which may be considered as an
additional constraints.

In the atlas ${\bf U}$ the integral~(\ref{action}) may be rewritten in the
following form~\cite{konst83n4}
\begin{eqnarray}
S=\sum\limits_{k\in P}\int\limits_{V_k}L(\Phi _A(x_k),\Phi _A,_\alpha
(x_k)) \sqrt{-g}d^4x_k - \nonumber  \\
\sum\limits_{k<l}\int\limits_{V_k\cap
V_l}L(\Phi _A(x_{kl}),\Phi _A,_\alpha (x_{kl}))\sqrt{-g}d^4x_{kl}+...
\nonumber \\
\label{modact}
+(-1)^{K}\sum\limits_{i_0<...<i_K}\int\limits_{V_{i_0}\cap ...\cap
V_{i_K}}L(\Phi _A(x_{i_0...i_K}),\Phi _A,_\alpha (x_{i_0...i_K}))\sqrt{-g}
d^{4}x_{i_0...i_K}
\end{eqnarray}
where $K<\infty $ because of standard supposition about paracompactness of
space-time manifold ${\bf M}$~\cite{hawkellis}.

For the following formalization of the action integral (1) consider the set $%
{\bf \Lambda }_{{\bf U}}$ of subsets of the set ${\bf P}$ such that $%
(i_0,...,i_k)\in {\bf \Lambda }_{{\bf U}}$ if and only if ${\bf V}_{i_0}\cap
...\cap {\bf V}_{i_k}\ \neq \emptyset $, where $\emptyset $ denotes the
empty set. The set ${\bf \Lambda }_{{\bf U}}$ which satisfies to such
condition is called a nerve of the covering ${\bf U}$ \cite{hilton} and its
elements of type $I_k=(i_0,...,i_k)$ are known as k-dimensional simplexes~%
\cite{hilton}. The zero-dimensional simplexes, i.e. elements of the type $%
I_0=i_0$, are vertexes. It is follows from definition that if $I_k\in {\bf %
\Lambda }_{{\bf U}}$ and $J_l\in I_k$, where $l<k$, then $J_l\in {\bf %
\Lambda }_{{\bf U}}$. This property shows that the nerve ${\bf \Lambda }_{%
{\bf U}}$ of the covering ${\bf U}$ is a particular case of the abstract
simplicial complex~\cite{hilton}. Such constructions are widely used in the
algebraic topology, in particular, in Cech cohomology theory whose
connections with topological quantization was discussed in~\cite{Alvarez}.

With the nerve ${\bf \Lambda }_{{\bf U}}$ of the covering ${\bf U}$ we may
associate the system of its characteristic functions which will be denoted
as ${\bf F}_{{\bf \Lambda }}=\left\{ f_{{\bf \Lambda }}^0,...,f_{{\bf %
\Lambda }}^K\right\} $ \cite{konst83n4}, \cite{padova1} where the functions $%
f_{{\bf \Lambda }}^l=f_{{\bf \Lambda }}^l(I_l)\in {\bf F}_{{\bf \Lambda }}$
, $0\leq l\leq K$, are defined on the set of natural numbers as follows
\begin{equation}
\label{charfunct}f_{{\bf \Lambda }}^l=\sum\limits_{i_0<...<i_l}a_{{\bf %
\Lambda }}^{I^l}f_{i_0}^0\wedge ...\wedge f_{i_l}^0
\end{equation}
where $i_m\in {\bf N}$, ${\bf N}$ is the set of natural numbers, ''$\wedge $%
'' denote exterior multiplication,
\begin{equation}
\label{f0}f_i^0=f_i^0(j)=\delta _{ij}
\end{equation}
and
\begin{equation}
\label{koef}a_{{\bf \Lambda }}^{I_l}=1,\ \ {\rm if}\ \ I_l\in {\bf \Lambda }%
_U,\ \ {\rm and}\ \ a_{{\bf \Lambda }}^{I_l}=0,\ \ {\rm if}\ \ I_l\notin
{\bf \Lambda }_{{\bf U}}
\end{equation}
It is follows from definitions that the connection of the system ${\bf F}_{%
{\bf \Lambda }}$ of characteristic function of nerve ${\bf \Lambda }_{{\bf U}%
}$ with atlas ${\bf U}$ is one-to one and hence ${\bf F}_{{\bf \Lambda }}$
defines the topology of the manifold ${\bf M}$ as well as corresponding
atlas ${\bf U}$.

Using the definitions~(\ref{charfunct})-(\ref{koef}) we may rewrite
equality~(\ref{modact}) as follows~\cite{konst83n4}, \cite{padova1}
\begin{equation}
\label{actff}S=\sum\limits_{k=0}^K(-1)^k\sum\limits_{i_0<...<i_k}S_k^{I_k}f_{%
{\bf \Lambda }}^k(I_k)
\end{equation}
where
\begin{equation}
\label{actcoef}S_k^{I_k}=\int\limits_{{\bf V}_{i_0}\cap ...\cap {\bf V}%
_{i_k}}L(\Phi _A(x_{i_0...i_k}),\Phi _A,_\alpha (x_{i_0...i_k}))\sqrt{-g}%
d^4x_{i_0...i_k}
\end{equation}

Equalities~(\ref{actff}), (\ref{actcoef}) together with definitions~(\ref{f0}%
)-(\ref{koef}) formalize the dependence of the action integral from the
topology of space-time. They make possible to do several observations.

First, equalities~(\ref{actff}) and (\ref{actcoef}) formalize the above
statement that the topology of manifold play the role of the additional
constraint. Second, the topology of arbitrary manifold may be coded by the
system ${\bf F}_{{\bf \Lambda }}$ which may be done finite for compact
manifolds. Some another methods of the manifold structure coding are
described in~\cite{fomenko}, but they are less suitable for our purpose. It
is known also, that independently from the method of the manifold structure
coding in three or more dimensions, the set of codes which describe all
manifolds of the given dimensionality is infinite with infinite subset of
codes which define the given manifold. Moreover, if dimension of manifold is
three or more then there is no so simple classification of manifold
structures as in two dimensions~\cite{fomenko}. Third, any changes in the
topology of manifold ${\bf M}$ may be represented as corresponding changes
of the system ${\bf F}_{{\bf \Lambda }}$. Really to change the topology of $%
{\bf M}$ it is necessary to change its atlas ${\bf U}$, i.e. the order in
which the coordinate maps ${\bf V}_i$ are joined with each other and their
number. The change of atlas ${\bf U}$ induce the change of its nerve ${\bf %
\Lambda }_{{\bf U}}$ and hence the system ${\bf F}_{{\bf \Lambda }}$ because
the correspondences ${\bf U}\leftrightarrow {\bf \Lambda }_{{\bf U}%
}\leftrightarrow {\bf F}_{{\bf \Lambda }}$ are one-to-one. However the
representation of the action functional $S$ in the form~(\ref{actff}) does
not contain any sign of the $F_{{\bf \Lambda }}$ changes. Moreover, such
representation does not contains any information about joining conditions in
the intersections of coordinate maps. So, any changes of space or space-time
topology may be done only ''by hands'' and does not follow from the general
formalism. Therefore the standard methods of the current field theory (both
classical and quantum in its path-integral form), which are based on the
usage of the action functional $S$, does not permit to describe the
dynamical change of space-time topology. To make possible such description
it is necessary to use functionals which contain not only ${\bf F}_{{\bf %
\Lambda }}$ but also some objects that may be called as ''discrete
derivatives'' of ${\bf F}_{{\bf \Lambda }}$ (as an example of such objects
may be used operators $\rho _{I_k}^{\pm }$ \cite{konst83n4}, \cite{padova1}
which may be interpreted as creation and annihilation operators of the
simplex $I_k$). The introduction of such objects is equivalent to
introduction of some non local (topological) interaction which has no
analogies in the current field theory therefore it is almost hopeless to
solve this problem directly but it is possible to investigate some
possibilities in the construction of the consecutive topology change theory
and its main features in the scope of the standard theories.

In the above the general 4-dimensional form of the action integral was
considered while in the context of the topology change description the usage
of some parameterization would be more suitable. The introduction of such
parameterization is straightforward and we do not consider it here. In
particular, in parameterized form the action integral~(\ref{action}) reads
\begin{equation}
\label{paramact}S=\int Ldt
\end{equation}
where the Lagrangian $L$ is defined analogously to~(\ref{actff}), (\ref
{actcoef}):
\begin{equation}
\label{paramlagr}L=\sum\limits_{k=0}^K(-1)^k\sum%
\limits_{i_0<...<i_k}L_k^{I_k}f_\Lambda ^k(I_k,t)
\end{equation}
where $f_{{\bf \Lambda }}^k(I_k,t)$ and $L_k^{I_k}$ are straightforward
analog of (4)-(6), (8).

It is easy to see that parameterization of the action integral does not
change result: (i) both in general four-dimensional form and in the
parameterized form the topological variables are the discrete valued
functions, and (ii) the action functional $S$ contains the topological
variables only as parameters (or constraints).

\section{Topology change in four-dimensional theory: application of Morse
theory}

To simplify the problem consider the particular case than the part of
space-time is a compact four-manifold ${\bf M}^4$ whose boundary is a
disjoint sum of three-dimensional space-like manifolds ${\bf M}_1^3$ and $%
{\bf M}_2^3$: i.e. $\partial {\bf M}^4={\bf M}_1^3\cup {\bf M}_2^3$, and $%
{\bf M}_1^3\cap {\bf M}_2^3=\emptyset $. The manifold ${\bf M}^4$ is often
called as interpolating manifold. Such models may be described in the
framework of Morse theory \cite{milnor}, \cite{wallece} which state that:

(i) there is a smooth function $\varphi $ on the manifold ${\bf M}^4$, such
that $0\leq \varphi (p)\leq 1$ for all $p\in {\bf M}^4$, $\varphi
(p_1)\equiv 0$ for all $p_1\in {\bf M}_1^3$, $\varphi (p_2)\equiv 1$ for all
$p_2\in {\bf M}_2^3$, and moreover $\varphi $ has a finite number of
nondegenerate critical points on ${\bf M}^4$ (the point $p\in {\bf M}^4$ is
called the nondegenerate critical point of smooth function $\varphi $ if in
arbitrary system of local coordinates $x^\alpha $ the following conditions
are satisfied: $\varphi ,_\alpha (p)=0$ and $\det \left\| \varphi ,_{\alpha
\beta }(p)\right\| \neq 0$);

(ii) ${\bf M}_2^3$ may be obtained from ${\bf M}_1^3$ by a finite number of
spherical modifications which correspond to the critical points of $\varphi$.

The correspondence between the non-degenerate critical points of function $%
\varphi $ and the topology of the level surfaces of this function is the
follows \cite{milnor}.

Let $p_{*}$ is a non-degenerate critical point of $\varphi $, $\varphi
(p_{*})=c$ and let there is no other critical points on the level surface $%
\varphi =c$. In some neighborhood of $p_{*}$ a system of local coordinates $%
\left\{ x^\alpha \right\} $ exists such that $\varphi $ is represented in
the form
\begin{equation}
\label{morsfunct}\varphi =\varphi (p_{*})+\frac 12\sum\limits_{\alpha
=1}^4a_\alpha \cdot (x^\alpha )^2
\end{equation}
where coefficients $a_\alpha $ are equal to $\pm 1$. Let $r+1$ is the number
of negative $a_\alpha $ in~(\ref{morsfunct}): $r+1=Ind_{-}\left\| \varphi
,_{\alpha \beta}(p_{*})\right\| $. Then, the level surface $\varphi
=c+\epsilon $ may be obtained from the level surface $\varphi =c-\epsilon $,
where $\epsilon=const>0$, through the spherical modification of rank $r$
\cite{milnor}, \cite{wallece}. In the case of 3-manifolds such modification
may be represented as a contraction of the sphere ${\bf S}^r$ into the
critical point $p_{*}$ along the trajectories of the vector field $l_\alpha
=\varphi,_\alpha $ and the following inflation of the sphere ${\bf S}%
^{3-r-1} $ from the same point $p_{*}$. Within this the contraction of ${\bf %
S}^r$ is realized in the subspace with coordinates $x^\alpha $ which
coincides with $a_\alpha =-1$ in~(\ref{morsfunct}), while the inflation of $%
{\bf S}^{3-r-1}$ is realized in the subspace with coordinates $x^\beta $
with $a_\beta =+1$. More formally such modification is described by the
equation
\begin{equation}
\label{sphermod}{\bf M}_2^3=\left( {\bf M}_1^3\setminus ({\bf E}^{3-r}\times
{\bf S}^r\right) \cup \left( {\bf E}^{r+1}\times {\bf S}^{3-r-1}\right)
\end{equation}
where ${\bf M}_1^3$ and ${\bf M}_2^3$ are the level surfaces of $\varphi
=c-\epsilon $ and $\varphi =c+\epsilon $ respectively and ${\bf E}%
^{3-k}\times {\bf S}^k$ is the tubular neighborhood of a directly embedded
sphere $S^k$. The generalization of such procedure on arbitrary-dimensional
case is straightforward.

The 3-sphere creation ($\emptyset \rightarrow {\bf S}^3$) and its
annihilation ( ${\bf S}^3\rightarrow \emptyset $) are described by spherical
modifications of rank $r=-1$ and $r=3$ respectively, while the wormhole
creation is the spherical modification of the zero rank.

It is obvious that the spherical modifications theory may be applied to the
description of 3-spaces topology changes on the given four-manifold ${\bf M}%
^4$ with given topological structure, because the topological structure of
manifold must be given before introduction of smoothness and before
definition of Morse function $\varphi $ (or arbitrary smooth function). The
main principles of such application are the follows (for details see~\cite
{konst83n12} - \cite{konst88} ).

It is supposed that the level surfaces $\varphi =$const are space-like and
the vector fields $l_\alpha =\varphi ,_\alpha $ is time-like everywhere
except the critical points of $\varphi $, where $l_\alpha =0$. Outside the
critical points of $\varphi $ the metric tensor of space-time may be
represented in the form
\begin{equation}
\label{metric}g_{\alpha \beta }=\frac{2l_\alpha l_\beta }f-\widetilde{g}%
_{\alpha \beta }
\end{equation}
where $f=g^{\rho \sigma }l_\rho l_\sigma =\widetilde{g}^{\rho \sigma }l_\rho
l_\sigma $ and $\widetilde{g}_{\alpha \beta }$ is a positive definite metric
on ${\bf M}^4$.

Using representation~(\ref{metric}) we may investigate the asymptotic
properties of space-time models in the vicinity of nondegenerate critical
point of function $\varphi $ (i.e. e. in the vicinity of the topology change
points). In particular, the direct calculation of Ricci tensor and scalar
curvature gives
\begin{equation}
\label{ricci}R_{\alpha \beta }=\widetilde{R}_{\alpha \beta
}+\sum\limits_{n=1}^4\left( \frac 1f\right) ^n\stackrel{n}{R}_{\alpha \beta }
\end{equation}
and
\begin{eqnarray}
R=-\widetilde{R}+\left( \frac 1f\right) \left\{ 4l^\alpha l^\beta
\widetilde{R}_{\alpha \beta }-2\left( (l_{\mid \alpha }^\alpha )^2-l_{\mid
\beta
}^\alpha l_{\mid \alpha }^\beta \right) -2f^\alpha ,_\alpha \right\} +
\nonumber  \\
\label{scalcurv}
\left( \frac 2f\right) ^2\left\{ 2l_{\mid \alpha }^\alpha l^\sigma f,_\sigma
+2f,^\alpha f,_\alpha +l^\alpha l^\beta f,_{\alpha \mid \beta }\right\}
-\frac 4{f^3}\left( l^\alpha f,_\alpha \right) ^2
\end{eqnarray}
where tilde ''\symbol{126}'' denote quantities correspond to the metric $%
\widetilde{g}_{\alpha \beta }$ and the $\stackrel{n}{R}_{\alpha \beta }$ are
certain polynomials on $l_\alpha $, $f,_\alpha $ and their covariant
derivatives with respect to $\widetilde{g}_{\alpha \beta }$ (The explicit
form of $R_{\beta \gamma \delta }^\alpha $ and $R_{\alpha \beta }$ are given
in~\cite{konstmel86} ).

It is easy to see from~(\ref{morsfunct}), (\ref{metric}) that non-degenerate
critical points of $\varphi $ are essentially singular points of metric $%
g_{\alpha \beta }$: the limits of right-hand side of~(\ref{metric}) at these
points exists, but depends on the direction~\cite{konst83n12}, \cite{konst85}%
. Further, it is easy to see that near non-degenerate critical points of $%
\varphi $ the curvature tensors of space-time have the following asymptotic
\begin{equation}
\label{asymptcoef}\stackrel{n}{R}_{\beta \gamma \delta }^\alpha ,\ \ \
\stackrel{n}{R}_{\alpha \beta }\sim f^{n-1}
\end{equation}
and hence
\begin{equation}
\label{asympt}R_{\beta \gamma \delta }^\alpha ,\quad R_{\alpha \beta },\quad
R\sim \frac 1f
\end{equation}
Therefore the topology change points are the scalar curvature singularities
of space-time. More detailed investigation of space-time properties near
topology change points and some simple examples may be found elsewhere~\cite
{konst85}, \cite{konstmel86}.

The representation of the Lorentzian metric $g_{\alpha \beta }$ in the form (%
\ref{metric}) may be used not only for investigation of space-time near the
critical points of the function $\varphi $ (i.e. near the topology change
points) but for the construction of some variants of the topology change
theory on the given four manifold. For this purpose the scalar function $%
\varphi $ and positive definite metric $\widetilde{g}_{\alpha \beta }$ are
used as new independent variables instead of the Lorentzian metric $%
g_{\alpha \beta }$ which defines the motion of the sources. Such program
leads to a new class of scalar-tensor theories of gravity, which partially
realized Hawking's idea about the Euclidean nature of space-time~\cite
{hawk78}. The outline of such model theories were discussed in~\cite{konst85}%
-\cite{konst88}. In particular, the action integral in such theory in the
approximation of minimal coupling with sources may be written in standard
form
\begin{equation}
\label{actst}L=\int \left( L_g+\kappa L_m\right) d\sigma
\end{equation}
where $L_g$ is the gravitational Lagrangian depending on scalar field $%
\varphi $ and metric tensor $\widetilde{g}_{\alpha \beta }$ and their
derivatives, $L_m$ is the standard Lagrangian of the source fields $\theta _A
$ ($A$ is the cumulative index) and their covariant derivatives with respect
to $g_{\alpha \beta }$, $d\sigma $ is invariant volume element. The
stationarity condition gives
\begin{eqnarray}
\delta L_g/\delta \widetilde{g}^{\alpha \beta }-\kappa T_{\rho \sigma
}D_{\alpha \beta }^{\rho \sigma }=0, \label{metreq} \\
\label{phieq}
\delta L_g/\delta \varphi -\kappa (T_{\rho \sigma }P^{\rho \sigma \alpha
})_{\mid \alpha }=0
\end{eqnarray}
and
\begin{equation}
\label{mattereq}\delta L_m/\delta \theta _A=0
\end{equation}
where $\delta /\delta \widetilde{g}^{\alpha \beta }$, $\delta /\delta
\varphi $ and $\delta /\delta \theta _A$ are variational derivatives, $%
T_{\alpha \beta }$ is the standard energy-momentum tensor of the source
fields (the same as in classical general relativity) and tensors $D_{\alpha
\beta }^{\rho \sigma }$ and $P^{\rho \sigma \alpha }$ are equal to
\begin{eqnarray}
D_{\alpha \beta }^{\rho \sigma }=\partial g^{\rho \sigma }/\partial
\widetilde{g}^{\alpha \beta }=\frac 4fl^{(\rho }\delta _{(\alpha }^{\sigma
)}l_{\beta )}-\frac 4{f^2}l^\rho l^\sigma l_\alpha l_\beta -\delta _{(\alpha
}^\rho \delta _{\beta )}^\sigma,  \label{D} \\
\label{P}
P^{\alpha \beta \sigma }=\partial g^{\alpha \beta }/\partial \varphi
,_\sigma =\left( 4/f\right) l^{(\alpha }g^{\beta )\sigma }-\left(
4/f^2\right) l^\alpha l^\beta l^\sigma
\end{eqnarray}

Equations (\ref{mattereq}) are the classical equations for the source fields
and the equations (\ref{metreq}) and (\ref{phieq}) define the scalar field $%
\varphi $, the positive definite metric $\widetilde{g}_{\alpha \beta }$ and
the pseudo-Riemannian (Lorentzian) structure of space-time through (\ref
{metric}).

Some problems of such approach, namely, the choice of the Lagrangian $L_g$
and the singularities problem were discussed in~\cite{konstmel86}, \cite
{konstmel88}, \cite{konst85}-\cite{konst88}. Here we want to point out
several main features of such approach.

First, the structure of four-manifold must be given. This condition is very
restrictive, because it automatically excludes from consideration a big
class of topological changes or a big class of histories (intermediate
states).

Second, the equation (\ref{phieq}) is an elliptic one. So, this approach
does not lead to the dynamical description of the topological change,
because minimum one independent variable is a subject of the boundary
problem. In application to the Universe evolution it means that both initial
and final states of the Universe are given. The physical meaning of such
problem, in particular, the nature of such boundary conditions, is very
unclear.

At last, the direct application of the Morse theory lead to the singular
space-time models and singular theory as it is follows from the above
consideration. Using this fact De Witt made conclusion about dynamical
suppressing of topology changes in quantum gravity~\cite{dewitt}. His
conclusion were reanalyzed in~\cite{revised}. Nevertheless, De Witt
conclusion cannot be considered as a general theorem because both in~\cite
{dewitt} and \cite{revised} only particular topology change model were
considered in the framework of standard general relativity without any
references to some theory of topology change. To obtain nonsingular theory
some type of regularization~\cite{konst88} or so-called Lorentz cobordism~%
\cite{yodziz} may be used. In the first case the additional boundary
conditions must be introduced but the possibility to include them into the
general formalism is not obvious. In the Lorentz cobordism case some
additional restrictions on the 4-manifold structure is necessary and vector
field $l_\alpha $ in (\ref{metric}) become nonintegrable. The resulting
model will describe transition from ''initial'' manifold ${\bf M}_1$ to the
''final'' manifold ${\bf M}_2$ and may be both singular and regular but
non-causal~\cite{geroch}. To obtain nonsingular model some additional
surgical operations may be necessary. The consideration of full set of
histories between ''initial'' and ''final'' states is impossible in both
methods.

\section{Topology change in multidimensional gravity}

The most of the current unified theories require space-time to be of more
then four dimensions. Independently from the reasons which lead to the
nonobservability of additional dimensions, the multidimensionality of
space-time gives several possibilities for the dynamical description of
topology change of 3-space. Here we shall discussed briefly two such
possibilities.

\subsection{Effective topology change via dynamical inflation-contraction}

In the standard dynamical dimensional reduction paradigm multidimensional
space-time supposed to have the topological structure of direct product of
real line (global time coordinate) and several topological spaces, i.e.
$$
{\bf M}={\bf R}\times {\bf M}_1\times {\bf M}_2\times ...\times {\bf M}_k
$$
such that one of space, for instance ${\bf M}_1$, has dimension 3, while
dimension of other spaces may be arbitrary. It is supposed also, that all
spaces ${\bf M}_i$, $i=1,$...,$k$, are compact. The model of space-time is
constructed by such a way that 3-manifold ${\bf M}_1$ (our universe) expands
from initial singular or Plank state while other manifolds contract from the
state with finite scales to the Plank scales.

It may be speculatively supposed that universe may pass through several
stages such that at each stage one 3-manifold expands and one 3-manifold
contracts while other manifolds remain in the state with Plank scales. As a
result at different stages of evolution universe will have different
effective topologies.

Unfortunately, this idea has two serious deficiencies. First, to describe
the wide class of the topological changes the total dimension of
multidimensional space-time in such scheme must be enormously big (infinite)
while there are no quantum field theories with total dimension of space-time
more then 26. Second, it is very difficult to find some natural mechanism
which may control the transitions between different stages of the universe
evolution in such scheme. Therefore, the paradigm of effective topology
change of universe via inflation-contraction must be considered today as
pure principal possibility.

\subsection{Topology change in the embedding models}

Analysis of the Morse theory gives some ideas about topology change
description in multidimensional theories. Namely, it is easy to see that the
Morse function $\varphi $ defines embedding of space-time $({\bf M}^4,g)$ in
5-dimensional topological space ${\bf V}^5={\bf M}^4\times {\bf R}$. As the
natural generalization, we may consider our space-time as submanifold of
some multidimensional manifold with given topology, in particular, as a
submanifold of the Euclidean space of the appropriate dimensionality $N$
which depends from the type of embedding. For instance, for arbitrary
four-dimensional manifold $N=8$, 10 and (10+7) for smooth, isometrical
Riemannian and isometrical pseudo-Riemannian (with arbitrary metric
signature) embedding respectively~\cite{bishop}, \cite{rohlin}. In
difference with standard Kaluza-Klein-type theories we do not demand nor the
existence of any dimensional reduction mechanism, nor the representation of
the whole manifold, namely ${\bf R}^N$, as a direct product ${\bf M}^n\times
{\bf V}^{N-n}$, where ${\bf M}^n$ and ${\bf V}^{N-n}$ are some smooth
manifolds. Instead this the existence of two type of observes and fields is
supposed: the external observers and fields which are defined in the whole $%
{\bf R}^N$, and the internal observers and fields which are defined on ${\bf %
M}^n$. The class of internal fields may include both the induced fields and
the surface distributed fields. The full theory must contains two parts: the
description of the embedding and the description of the interior dynamics.

It is necessary to note, that the possibility of the space-time
consideration as a membrane in higher dimensional space is not new and were
discussed by several authors~\cite{pavsic}-\cite{gibbmembr}, but only the
local properties of embedding were considered. The most principle feature of
our approach is the consideration of both global and local properties of
embedding in the unified formalism.

Here we shall briefly discuss only the outline of such approach. The
detailed consideration is the subject of separate papers and will be
published elsewhere.

We begin our consideration from reproduction of some well-known facts about
embedding of manifolds into Euclidean space.

Let ${\bf M}^n$ is a smooth $n-$dimensional submanifold of the Euclidean
space ${\bf R}^N$ with Cartesian coordinates $\left\{ X^P{\rm ,}%
P=1,...,N\right\} $, and the number $k=N-n$ is called the codimension of $%
{\bf M}^n$. The embedding ${\bf M}^n\longrightarrow {\bf R}^N$ is defined
locally by the set of equations
\begin{equation}
\label{embloc}X^A=X^A\left( x^1,...,x^n\right)
\end{equation}
where $\left( x^1,...,x^n\right) $ are the local coordinates in ${\bf M}^n$
and the matrix $\left\| \partial X^P/\partial x^i\right\| $, $i=1,...,n$,
has maximal rank $n$. Equations~(\ref{embloc}) may be obtained as a solution
of the well known Gauss-Kodazzi equations~\cite{kobayashi}, \cite{maia}
which connect the intrinsic and extrinsic geometries of submanifolds.
Unfortunately, such description does not admit to define the global
structure of ${\bf M}^n$ or investigate its dynamics. Therefore we shell use
the alternative description of ${\bf M}^n$ as the intersection of $N-n$
hypersurfaces in ${\bf R}^N$, i.e. by the system of equations
\begin{equation}
\label{embglob}\Phi ^A\left( X^P\right) =0
\end{equation}
where $\Phi ^A\left( X^P\right) $, $A=1,...,k$, are some smooth functions on
${\bf R}^N$, and the matrix $\left\| \Phi ^A,_P\right\| =\left\| \partial
\Phi ^A/\partial X^P\right\| $ must have maximal rank, i.e.
\begin{equation}
\label{consist}rank\left\| \Phi ^A,_P\right\| =\left\| \partial \Phi
^A/\partial X^P\right\| =k=N-n
\end{equation}

To introduce the atlas on ${\bf M}^n$ let's consider arbitrary point $p\in
{\bf M}^n\subset {\bf R}^N$. By force of the implicit function theorem there
is a neighborhood ${\bf U}_p\subset {\bf R}^N$ of arbitrary point $p\in {\bf %
M}^n\subset {\bf R}^N$ such that the system (\ref{embglob}) may be solved in
the form (\ref{embloc}) with $x^i=X^{m_i}$, $i=1,...,n$ and the numbers $%
\left\{ m_1,...,m_n\right\} $ are a subset of the set $\left\{
1,...,N\right\} $. It means that there is a one-to-one correspondence
between the points of ${\bf U}_p\cap {\bf M}^n$ and one of $n$ -dimensional
coordinate planes in ${\bf R}^N.$ Taking different points of $p\in {\bf M}^n$
we may obtain the covering of ${\bf M}^n$ by the regions $\widetilde{{\bf U}}%
_p={\bf U}_p\cap {\bf M}^n$ which define the atlas on ${\bf M}^n$.

Equations (\ref{embglob}) are the algebraic constraints, but the smooth
functions $\Phi _A$ may be arbitrary. In particular, these fields may be
considered as usual scalar fields in ${\bf R}^N$. In the simplest model the
action functional will have the form
\begin{equation}
\label{extact}S_N=\int d^NX\left( \eta ^{PQ}\delta _{AB}\Phi ^A,_P\Phi
^B,_Q-V\left( \Phi ^A\right) \right)
\end{equation}
where $d^NX$ - is an invariant volume element of ${\bf R}^N$, $\delta _{AB}$
- is Kronecker symbols, $A,B=1,...,k$; $k=N-n$ is a codimension of ${\bf M}%
^n $, $\eta _{PQ}$ is the Euclidean or pseudo-Euclidean metric on ${\bf R}^N$
with given signature, $\Phi ^A,_P=\partial \Phi ^A/\partial X^P$ and $%
V\left( \Phi ^A\right) $ is the potential. Of cause, the action $S_N$ may
contain not only the scalar fields $\Phi ^A$ but additional scalar, vector
and tensor fields also. Moreover, instead of Kronecker symbols $\delta _{AB}$
an arbitrary nondegenerate $k\times k$ matrix may be used.

The set of equations
\begin{equation}
\label{top}\delta S_N\prime \delta \Phi _A=0
\end{equation}
together with the constraints (\ref{embglob}), the consistency condition (%
\ref{consist}) and appropriate initial or boundary conditions give dynamical
description of the topology of ${\bf M}^n$ by means of its embedding into $%
{\bf R}^N$. Such description of the topology of ${\bf M}^n$ does not unique
because the same manifold may be embedded into ${\bf R}^N$ by different
manners. The interior dynamics of fields on ${\bf M}^n$ does not described
by these equations.

To describe the dynamics of fields on $M^n$ the full action must contain
additional surface term. For definiteness we shall write this term in the
form
\begin{equation}
\label{intact}S_n=\int\limits_{\Phi _A=0}d^n\sigma \left( R_n+L_m\right)
\end{equation}
where $d^n\sigma $ - is an invariant volume element on ${\bf M}^n$, $R_n$ -
is a scalar curvature and $L_m$ - is the Lagrangian density of the matter
fields on ${\bf M}^n$. In general, $L_m$ may contain both induced fields, in
particular, the fields $X^A=X^A\left( x^1,...,x^n\right) $, and the fields
which are distributed on ${\bf M}^n$ (surface-distributed fields). In the
simplest case
\begin{equation}
\label{simplagr}L_m=\eta _{AB}g^{\mu \nu }X^A,_\mu X^B,_\nu
\end{equation}
where $g_{\mu \nu }$ is some metric on ${\bf M}^n$, and the action $S_n$
become the direct generalization of the Nambu-Goto string action. The full
action
\begin{equation}
\label{fullact}S=S_N+S_n
\end{equation}
describes both the topology of ${\bf M}^n$ (by means of its embedding into $%
{\bf R}^N$) and the dynamics of fields on ${\bf M}^n$. Equations
\begin{equation}
\label{einst}\delta S_n\prime \delta g^{\mu \nu }=0,
\end{equation}
\begin{equation}
\label{fe}\delta S_n\prime \delta X^A=0
\end{equation}
and other analogous equations describe the dynamics of fields on ${\bf M}^n$
and its internal geometry. It is easy to see that these equations give
additional constraints for the equations (\ref{embglob}) which define the
topology of ${\bf M}^n$ and the embedding ${\bf M}^n\longrightarrow {\bf R}^N
$.

Such scheme may be directly applied to the description of our space-time (in
particular, to the Universe evolution) if we put in the above equalities $%
n=4 $ and $N\geq 17$ (so that $k=N-n\geq 13)$. It is not contradict also to
the standard Kaluza-Klein approach, if we suppose that the multidimensional
space-time of the Kaluza-Kline theory is a submanifold of some ${\bf R}^N$.
However we will not consider such possibility here.

More detailed consideration of the considered approach is the subject of the
separate paper. Nevertheless some additional remarks are necessary. The
existence theorems gives only necessary conditions for isometrical
pseudo-Riemannian embedding. Unfortunately, the induced metric is not
necessary Lorentzian even if the space ${\bf R}^N$ is Lorentzian. To obtain
Lorentzian metric on ${\bf M}^n$ we may use several possibilities. First, we
may omit the isometrical condition and suppose that $g_{\mu \nu }$ is an
arbitrary metric with Lorentzian signature on $M^n$ (simultaneously the
necessary dimension of ${\bf R}^N$ will be reduced). Second we may demand
that induced metric $g_{\mu \nu }$ on ${\bf M}^n$ is Lorentzian. Both ways
are seems to be unsatisfactory because in the first case the connection
between external and internal geometries is very weak and in the second case
additional constraints on the functions $\Phi ^A\left( X^P\right) $ must be
introduced. It seems that the most appropriate choice is to introduce
additional smooth function $\Psi $ on ${\bf R}^N$ and take metric $g_{\mu
\nu }$ on ${\bf M}^n$ in the form (\ref{metric}), i.e.
\begin{equation}
\label{metr1}g_{\alpha \beta }=\frac{2l_\alpha l_\beta }f-\widetilde{g}%
_{\alpha \beta }
\end{equation}
where $l_\alpha $ and $f$ are the same as in (\ref{metric}), $\phi $ is the
restriction of $\Psi $ on ${\bf M}^n$, i.e. $\phi =\Psi \mid _M$, and $%
\widetilde{g}_{\alpha \beta }$ is induced by the Euclidean metric of ${\bf R}%
^N$. As an example of the function $\Psi $ the projection of ${\bf M}^n$ on $%
T$-axes may be used. It is easy to see, that if external space ${\bf R}^N$
has the only time-like direction then such projection will be defined by the
elliptic equation and so the definition of global structure of space-time
could not be reduced to the pure dynamical problem. Moreover, definition of
the metric of ${\bf M}^n$ in the above form to the same singularity problem
as in the case of direct application of Morse theory discussed in section 2.
To avoid this problem the condition of integrability of vector field $%
l_\alpha $ may be omitted. In this case the Lorentz cobordism models will be
included in the general formalism also.

\section{Conclusion}

We have consider some possible approaches to the description of the topology
of space-time and the topology changes in the framework of both
four-dimensional and multidimensional theories. Our results may be
summarized as follows.

First, the simplicial approach make possible to formalize the statement that
the topology of manifold play the role of the additional constraint. Namely,
in this approach topology of space or space-time is represented by means of
the system ${\bf F}_{{\bf \Lambda }}$ of characteristic function of the
nerve of some atlas of space-time. However action functional $S$ contains $%
{\bf F}_{{\bf \Lambda }}$ by linear manner and does not contain any sign of
the ${\bf F}_{{\bf \Lambda }}$ changes. Therefore the standard methods of
the current field theory (both classical and quantum in its path-integral
form), which are based on the usage of the action functional $S$, does not
permit to describe the dynamical change of space-time topology. To make
possible such description it is necessary to use functionals which contain
not only ${\bf F}_{{\bf \Lambda }}$ but also some objects that may be called
as ''discrete derivatives'' of ${\bf F}_{{\bf \Lambda }}$ (as an example of
such objects may be used operators $\rho _{I_k}^{\pm }$ \cite{konst83n4},
\cite{padova1} which may be interpreted as creation and annihilation
operators of the simplex $I_k$). The introduction of such objects is
equivalent to introduction of some non local (''topological'') interaction
which has no analogies in the current field theory.

Second, the direct application of Morse theory or its nonintegrable
generalization (Lorentz cobordism) make possible to describe topology change
on the given four-dimensional manifold. This condition is very restrictive
because they do not permit to consider all possible intermediate states or
all topological histories. Moreover, this approach leads to the singular
space-time models. To obtain nonsingular models some additional conditions
must be imposed or additional topological transformations must be made. Both
way are unsatisfactory because they could not be included in the general
formalism. Furthermore, the equation (\ref{phieq}), which defines the
simultaneity hypersurface is an elliptic one. Therefore the space topology
in such models is a subject of the boundary problem with boundary conditions
on different space-like hypersurfaces. The physical meaning of such boundary
conditions is unclear.

Third, the multidimensional theories give several possibilities for the
topology change description. The most radical way is to describe space-time
as a membrane in the Euclidean space of appropriate dimensionality. In
difference with the existing embedding space-time theories, the full action
in such approach must contain two terms: the term which define embedding of
4-dimensional space-time into Euclidean space of appropriate dimensionality
and the surface term which describe the dynamics of fields. Such approach
make possible to consider all topological histories including 4-manifolds
with exotic smoothness, whose possible role in physics was discussed
recently by Brans~\cite{brans}. Of cause, this approach does not free from
the number of difficulties most of which are the subject of separate paper.
Here we pointed out only two of such problems. The first one is pure
technical: it is also hopeless to obtain exact solution with non trivial
topology of space-time in any variant of such theory while any known
solution may be easily rewritten in such scheme. The second problem is of
principle nature because the finite classifications of smooth manifolds
whose dimension more or equal 3 does not exist. Moreover, the problem of
identification of manifolds is algorithmically unsolvable in 4 or more
dimensions and its algorithmic solvability in 3 dimensions is an open
question~\cite{fomenko}. Nevertheless, such approach is seems to be of
considerable interest because it is consistent with multidimensional
paradigm of the most current field theories and admit to consider all
possible topological configurations of 4-dimensional space-time.

\end{document}